\documentclass[twocolumn,showpacs,preprintnumbers,amsmath,amssymb,APSl,prd,nofootinbib,superscriptaddress]{revtex4-1}

\usepackage{dcolumn}
\usepackage{bm}
\usepackage{ifpdf}
\usepackage{hyperref}
\usepackage{bm}
\usepackage{xcolor,color,graphicx,graphics}
\usepackage[spanish,english]{babel}
\usepackage[latin1]{inputenc}
\usepackage[OT1]{fontenc}
\usepackage{latexsym,amssymb,amsmath,amsfonts}
\usepackage{makeidx}
\usepackage{epsfig,subfigure}
\usepackage{natbib}
\usepackage{epstopdf}
\usepackage{mathrsfs}
\usepackage{hyperref}
\hypersetup{colorlinks=true, linkcolor=blue, citecolor=green}
\usepackage{enumerate}

\definecolor{red}{rgb}{1,0,0}

\def\+{^\dagger}

\def\<{\leftarrow}
\def\>{\rightarrow}

\def\({\left(}
\def\){\right)}

\def\m{\mu} \def\n{\nu}    
\def\k{\kappa}\def\S{\Sigma}

\def\q{\quad}\def\qq{\qquad}

\newcommand{\bi}{\begin{itemize}} 				\newcommand{\ei}{\end{itemize}}
\newcommand{\benu}{\begin{enumerate}} 		\newcommand{\enu}{\end{enumerate}}
\newcommand{\bd}{\begin{dinglist}{0}}     \newcommand{\ed}{\end{dinglist}}
\newcommand{\bfig}{\begin{figure}[htbp]}  \newcommand{\efig}{\end{figure}}
        			
\newcommand{\bc}{\begin{center}} 				  \newcommand{\ec}{\end{center}}
\newcommand{\be}{\begin{equation}} 				\newcommand{\ee}{\end{equation}}
\newcommand{\bsub}{\begin{subequations}}  \newcommand{\esub}{\end{subequations}}
\newcommand{\ben}{\begin{eqnarray}} 			\newcommand{\een}{\end{eqnarray}}
\newcommand{\ba}[1]{\begin{array}{#1}} 		\newcommand{\ea}{\end{array}}
\newcommand{\bea}{\begin{equation}\begin{array}{rcl}}
\newcommand{\eea}{\end{array}\end{equation}}

\begin{document}
\title{An infinite class of exact rotating black hole metrics of modified gravity}

\author{Victor I. Afonso} \email{viafonso@df.ufcg.edu.br}
\affiliation{Unidade Acad\^{e}mica de F\'{\i}sica, Universidade Federal de Campina
    Grande, 58.429-900 Campina Grande-PB, Brazil}
\affiliation{Departamento de F\'{\i}sica, Universidad del Pa\'{\i}s Vasco UPV/EHU, Leioa-48940, Spain}
	
\author{Gerardo Mora-P\'{e}rez} \email{moge@alumni.uv.es}
\affiliation{Departamento de F\'{i}sica Te\'{o}rica and IFIC, Centro Mixto Universidad de Valencia - CSIC.
	Universidad de Valencia, Burjassot-46100, Valencia, Spain}

\author{Gonzalo J. Olmo} \email{gonzalo.olmo@uv.es}
\affiliation{Departamento de F\'{i}sica Te\'{o}rica and IFIC, Centro Mixto Universidad de Valencia - CSIC.
	Universidad de Valencia, Burjassot-46100, Valencia, Spain}
	
	\author{Emanuele Orazi} \email{orazi.emanuele@gmail.com}
\affiliation{ International Institute of Physics, Federal University of Rio Grande do Norte,
Campus Universit\'ario-Lagoa Nova, Natal-RN 59078-970, Brazil}
\affiliation{Escola de Ciencia e Tecnologia, Universidade Federal do Rio Grande do Norte, Caixa Postal 1524, Natal-RN 59078-970, Brazil}

\author{Diego Rubiera-Garcia} \email{drubiera@ucm.es}
\affiliation{Departamento de F\'isica Te\'orica and IPARCOS, 	Universidad Complutense de Madrid, E-28040 Madrid, Spain}

\date{\today}
\begin{abstract}
We build an infinite class of exact axisymmetric solutions of a metric-affine gravity theory, namely, Eddington-inspired Born-Infeld gravity, coupled to an anisotropic fluid as a matter source. The solution-generating method employed is not unique of this theory but can be extended to other Ricci-Based Gravity theories (RBGs), a class of theories built out of contractions of the Ricci tensor with the metric. This method exploits a correspondence between the space of solutions of General Relativity and that of RBGs, and is independent of the symmetries of the problem. For the particular case in which the fluid is identified with non-linear electromagnetic fields we explicitly derive the corresponding axisymmetric solutions. Finally, we use this result to work out the counterpart of the Kerr-Newman black hole when Maxwell electrodynamics is set on the metric-affine side. Our results open up an exciting new avenue for testing new gravitational phenomenology in the fields of gravitational waves and shadows out of rotating black holes.
\end{abstract}

\maketitle

\section{Introduction}

We are currently witnessing the beginning of a new era in gravitational physics ignited by the consolidation of multimessenger astronomy, namely, astronomy with different carriers: electromagnetic radiation (now including shadows), gravitational waves, neutrinos and cosmic rays. The fields of gravitational waves and shadows have actually reached their adulthood very recently. Indeed, the detection of gravitational waves out of binary mergers \cite{LIGOScientific:2016aoc,LIGOScientific:2017vwq}, and the outline of the bright ring of radiation induced by the super-heated plasma surrounding the supermassive central object of the M87 galaxy \cite{EventHorizonTelescope:2019dse}, are now powerful tools at disposal of the whole community working on the determination of the nature of compact objects, and on revealing the intimate nature of the gravitational interaction on its - yet mostly untested - strong field regime \cite{Berti:2015itd}.

Within the enormous theoretical pool of compact objects available to run tests with \cite{Cardoso:2019rvt}, black holes are undoubtedly still the privileged candidate. The uniqueness theorems \cite{Heusler}, our understanding of gravitational collapse \cite{JoshiBook}, and the electromagnetic phenomenology tested so far \cite{Bambi:2015kza}, singles out the Kerr(-Newman) family of solutions, described by mass, angular momentum and charge (the latter being typically neglected in astrophysical environments, see however \cite{Zajacek:2019kla}) as the embodiment of the black hole paradigm within General Relativity (GR). When looking for theoretical alternatives to it, particularly within gravitational extensions of GR, most attempts typically  assume slow-rotation motion \cite{Pani:2011gy,Yagi:2012ya,Barausse:2012qh,Ayzenberg:2014aka,Maselli:2015yva,Adair:2020vso}, either in order to decrease the degree of difficulty in solving the field equations of the theory of gravity under consideration or to implement numerical recipes, though fully-rotating solutions are also known \cite{Cembranos:2011sr, Ding:2019mal, Cano:2019ore, Jusufi:2019caq}.

The main aim of this work is to use the theoretical weaponry developed in the last few years, targeted to crack the structure of the field equations of a certain family of metric-affine gravity theories ({\it i.e.} with independent metric and affine connection), to obtain an infinite class of exact axisymmetric solutions within it. Such a family  is built via scalars out of contractions of the metric with the Ricci-tensor of the affine connection. These are  dubbed as {\it Ricci-based gravities} (RBGs), and the main weapon is the so-called {\it mapping method} \cite{Afonso:2018bpv}. This method establishes a correspondence between the spaces of solutions of two RBG theories, in particular allowing to generate solutions of a given RBG starting from a seed GR solution, via {\it purely algebraic transformations} \cite{Afonso:2019fzv,Olmo:2020fnk,Guerrero:2021avm}. Critical in this procedure is the identification of the matter Lagrangians sourcing the RBG/GR sides of the map, a point that has been systematically implemented for several types of matter fields \cite{Afonso:2018hyj,Delhom:2019zrb}. In the present work we shall consider the case of anisotropic fluids as the matter source, while our target RBG theory will be given by the so-called Eddington-inspired Born-Infeld gravity (EiBI), a suitable choice by its good algebraic properties as well as by its many applications in astrophysics and cosmology \cite{BeltranJimenez:2017doy}. By considering the case of non-linear electrodynamics formulated as anisotropic fluids, our main result will be the finding of the counterpart of the Kerr-Newman solution in EiBI gravity coupled to a Maxwell field. The potential applications of this novel result in the fields of gravitational waves and shadows will also be discussed.

\section{The two frames of RBGs and the mapping method}

RBGs are built as contractions of the space-time metric $g_{\mu\nu}$ with the (symmetric) part of the Ricci tensor of an independent  connection, $R_{\mu\nu}(\Gamma)$, via traces of powers of the object ${M^\mu}_{\nu} \equiv g^{\mu\alpha}R_{\alpha\nu}$. The symmetric Ricci requirement is (mostly) imposed in order to guarantee projective invariance, which safeguards the theory against ghost-like instabilities and trivialises the role of torsion  \cite{Afonso:2017bxr,BeltranJimenez:2019acz,BeltranJimenez:2020sqf}. The corresponding action is written as
\begin{eqnarray} \label{eq:actRBG}
\mathcal{S}_{RBG}&=& \frac{1}{2\kappa^2} \int d^4x \sqrt{-\text{det}(g)} \mathcal{L}_G(g_{\mu\nu},R_{(\mu\nu)}(\Gamma)) \nonumber \\
&+& \mathcal{S}_m(g_{\mu\nu},\psi_m)  \ ,
\end{eqnarray}
where $\kappa^2$ is Newton's constant in suitable units, the gravitational Lagrangian $\mathcal{L}_G$ is built with scalars out of the ${M^\mu}_{\nu}$ object above, while the matter sector implements a minimal coupling with a set of matter fields $\psi_m$ via $\mathcal{S}_m=\int d^4x \sqrt{-\text{det}(g)} \mathcal{L}_m(g_{\mu\nu},\psi_m)$. The field equations for the action (\ref{eq:actRBG}) are obtained by independent variation with respect to the metric and the connection (metric-affine or Palatini formalism), and can be conveniently rewritten in an Einstein-frame representation of the form (see \cite{Afonso:2018bpv} for details)
\begin{equation} \label{eq:Einfra}
{G^\mu}_{\nu}(q)=\tilde{T}{^\mu}_{\nu}(q) \ ,
\end{equation}
via the introduction of a new metric $q_{\mu\nu}$, which is compatible with the independent connection ({\it i.e.} $\Gamma$ is Levi-Civita of $q$), and is related to the space-time metric through the algebraic relation
\begin{equation}\label{eq:ommat}
q_{\mu\nu}=g_{\mu\alpha}{\Omega^\alpha}_{\nu} \ ,
\end{equation}
where the {\it deformation matrix\,}  ${\Omega^\mu}_{\nu}$ depends on the RBG theory chosen, but can always be written on-shell as a function of the matter fields (and likely the space-time metric $g_{\mu\nu}$ too). The energy-momentum tensor  appearing in (\ref{eq:Einfra}) is given by
\begin{equation} \label{eq:Ttotilde}
\tilde{T}{^\mu}_{\nu}(q)=\frac{1}{\vert \Omega \vert^{1/2}}  \left({T^\mu}_{\nu}(g)-\delta^{\mu}_{\nu} \left(\mathcal{L}_G(g,\psi_m)+\frac{T(g)}{2}\right)\right)
\end{equation}
where $T_{\mu\nu}(g)=\tfrac{-2}{\sqrt{-\text{det}(g)}} \tfrac{\delta \mathcal{S}_m}{\delta g^{\mu\nu}}$ is the usual energy-momentum tensor, while vertical bars denote a determinant.  Provided that all $g$-dependencies in the right-hand side of Eq.(\ref{eq:Ttotilde}) can be systematically replaced by their $q$-counterparts using (\ref{eq:ommat}), this provides not only a relation between the matter and gravitational sectors of two RBG theories, but also between their spaces of solutions. Since GR belongs to the RBG family via the trivial choice $\mathcal{L}_G = \text{tr}({M^\mu}_{\nu})$, the main attractiveness of this procedure lies on the possibility of mapping known solutions of GR into any other RBG of interest.

For the sake of this work let us implement the idea above for the case of anisotropic fluid matter sources, which covers many cases of physical interest, such as (obviously) perfect fluids as well as scalar and electromagnetic fields. Its energy-momentum tensor (in the RBG frame) is written as
\be \label{eq:TmnRBG}
{T^\mu}_{\nu}(g)=(\rho+p_{\perp})u^\mu u_\nu + p_{\perp}\delta^{\mu}_{\nu} +(p_r-p_{\perp})\chi^\mu \chi_\nu \ ,
\ee
with (mutually orthogonal) unit time-like, $g^{\mu\nu}u_{\mu}u_{\nu}=-1$, and space-like, $g^{\mu\nu}\chi_{\mu}\chi_{\nu}=1$, vectors while $\{\rho, p_r, p_{\perp}\}$ are the energy density, radial pressure, and tangential pressure, respectively.  Similarly, one can introduce a formally analogous energy-momentum tensor in the Einstein frame as
\be \label{eq:TmnGR}
\tilde{T}{^\mu}_{\nu}(q)=(\rho^q+p_{\perp}^q)v^\mu v_\nu + p^q_{\perp}\delta^{\mu}_{\nu} +(p_r^q -p_{\perp}^q)\xi^\mu \xi_\nu \ ,
\ee
with new  time-like, $q^{\mu\nu}v_{\mu}v_{\nu}=-1$, and space-like, $q^{\mu\nu}\xi_{\mu}\xi_{\nu}=1$, vectors and fluid's functions $\{\rho^q, p^q_r, p^q_{\perp}\}$. Using Eq.(\ref{eq:Ttotilde})
and the identifications $u^{\m}u_{\n}=v^{\m}v_{\n}$ and $\chi^{\m}\chi_{\n}=\xi^{\m}\xi_{\n}$,
we find the relations between the functions on each frame as \cite{Afonso:2018bpv}
\begin{eqnarray}
p^q_{\perp}&=&\frac{1}{\vert \Omega \vert^{1/2}} \left( \frac{\rho-p_r}{2} - \mathcal{L}_G\right) \label{eq:mapp1}  \\
\rho^q+p_{\perp}^q&=& \frac{(\rho+p_{\perp})}{\vert \Omega \vert^{1/2}} \label{eq:mapp2}\\
p^q_r-p^q_{\perp}&=& \frac{(p_r-p_{\perp})}{\vert \Omega \vert^{1/2}} \ . \label{eq:mapp3}
\end{eqnarray}
On the other hand, the deformation matrix itself can be written as an infinite power series of the energy-momentum tensor (see \cite{Jimenez:2021sqd} for details), which in the present case can be recast in terms of the fluid's functions (in the Einstein frame) as ${\Omega^\mu}_{\nu}=\alpha\, \delta^{\mu}_{\nu} + \beta\, v^{\mu}v_{\nu}+\gamma\, \xi^{\mu}\xi_{\nu}$, with $\{\alpha,\beta,\gamma\}$ being model-dependent functions.
Therefore, the mapping equations (\ref{eq:mapp1})-(\ref{eq:mapp3}) provide a relation between the functions characterizing the fluid on each frame. This allows one to reconstruct the corresponding matter Lagrangians yielding them, while the fundamental relation (\ref{eq:ommat}) maps the corresponding metric solutions of each frame. Thus, once we choose a specific RBG theory and some matter field content represented by an anisotropic fluid, this procedure allows us to map any specific solution of interest of such an RBG to a solution of any other RBG, in particular, GR itself. This is so because the mapping holds true irrespective of any symmetries of the background metric; in other words, it maps the entire spaces of solutions of GR+$\tilde{\mathcal{L}}_m(\psi_m,q)$ and a given RBG+$\mathcal{L}_m(\psi_m,g)$.

Among the pool of RBG theories, the so-called Eddington-inspired Born-Infeld gravity (EiBI) is particularly amiable for calculations.  Indeed, in this case the shape of the deformation matrix is given by the simple algebraic relation (see \cite{BeltranJimenez:2017doy} for details)
\begin{equation} \label{eq:OmEiBI}
\vert \Omega \vert^{1/2} {(\Omega^{-1})^\mu}_{\nu}=\lambda \delta^{\mu}_{\nu}-\epsilon \kappa^2 {T^\mu}_{\nu} \ ,
\end{equation}
where $\lambda$ is a constant related to the asymptotic character of the solutions (from now on we set $\lambda=1$ for asymptotic flatness), while the length-squared parameter $\epsilon$ encodes the deviations from GR, such that the EiBI Lagrangian density, which can be expressed as $\mathcal{L}_G=\tfrac{\vert \Omega \vert^{1/2}-1}{\epsilon \kappa^2}$, recovers the Einstein-Hilbert one in the limit $\vert R_{\mu\nu} \vert \ll \epsilon^{-1}$.
Combining (\ref{eq:OmEiBI}) with the set of mapping equations (\ref{eq:mapp1})-(\ref{eq:mapp3}), and using the energy-momentum tensor (\ref{eq:TmnRBG}), the fundamental relation \eqref{eq:ommat} allows us to write the space-time metric under the compact form
\begin{eqnarray}
g_{\mu\nu}&=&\left(1-\tfrac{\epsilon \kappa^2(\rho^q-p_r^q)}{2}\right)q_{\mu\nu} \nonumber \\
&&\;-\epsilon \kappa^2 \left[(\rho^q+p_{\perp}^q)v_\mu v_\nu + (p_r^q-p_{\perp}^q)\xi_\mu \xi_\nu\right] \ , \q \label{eq:ggen}
\end{eqnarray}
where the right-hand side only involves variables of the Einstein frame solution.
Therefore, given a seed solution obtained within GR, {\it i.e.}, a $q_{\mu\nu}$ metric plus a fluid configuration $\{\rho^q,p_r^q,p_{\perp}^q\}$ supporting it, the above equation provides the corresponding RBG solution. In what follows we shall particularize this formalism to the case of axially symmetric solutions.

\section{Rotating black holes}

Consider the general parametrization of a static, spherically symmetric metric as
\begin{equation}\label{eq:sss}
ds_q^2=-f(r)dt^2+\frac{dr^2}{f(r)}+h(r)d\Omega^2 \ ,
\end{equation}
where $d\Omega^2=d\theta^2+r^2\sin^2  \theta \, d\phi^2$. For the sake of the present work we shall fix the radial function as $h(r)=r^2$, so that the line element above, assumed to arise as a solution of GR coupled to some matter source, is characterized by a single function $f(r)$. In several works of recent years, the Janis-Newman algorithm (see \cite{Erbin:2016lzq} for a review) has been employed to arrive at the axisymmetric counterpart of (\ref{eq:sss}). In Boyer-Lindquist coordinates, its line element is written as  (see \cite{Azreg-Ainou:2014pra,Azreg-Ainou:2014aqa,Toshmatov:2017zpr} for details on this solution)
\begin{eqnarray}
ds_q^2&=&-\left(1-\frac{2\eta r}{\Sigma}\right)dt^2+\frac{\Sigma}{\Delta}dr^2 - 2a \sin^2 \theta \frac{2\eta r}{\Sigma} dt d\phi \nonumber \\
&+&  \Sigma\, d\theta^2  + \frac{\sin^2 \theta}{\Sigma} \left[(r^2+a^2)^2-a^2 \Delta \sin^2 \theta)\right] d\phi^2 \qq\ , \label{eq:rotmetric}
\end{eqnarray}
with the following definitions:
\begin{eqnarray}
\Sigma&=&r^2+a^2 \cos^2 \theta, \\
2\eta&=&r(1-f), \label{eq:eta} \\
\Delta&=&r^2f+a^2=r^2-2\eta r +a^2 \ ,
\end{eqnarray}
where $a=J/M$ is the spin parameter, with $0<a\leq 1$ and bounded from above by the extremality condition $M=J$. The Kerr-Newman solution belongs to this family under the choice $f(r)=1-\tfrac{2M}{r}+ \tfrac{Q^2}{r^2}$, with $M$ the ADM mass of the space-time and $Q$ the electric charge.

Due to its rather general character, the line element (\ref{eq:rotmetric}) will be our seed metric $q_{\mu\nu}$ in order to generate our infinite set of axisymmetric solutions of EiBI gravity. Assuming motion in the $\phi$ direction, the unit time-like vector of the fluid sourcing (\ref{eq:rotmetric}) can be written as $v^{\mu}=(v^t,0,0,v^{\phi})$, with the angular part typically written as $v^{\phi}=\omega v^t$, where $\omega \equiv d\phi/dt$ is the angular velocity of the fluid. In order to fix these functions $v^t$ and $\omega$, we introduce an orthornormal basis $e_{\alpha}^{(\mu)}$ given by the so-called Carter tetrad, whose components take the form \cite{Carter}
\begin{eqnarray}
\q e_{t}^{(\mu)}&=&\tfrac{1}{\sqrt{\Sigma \Delta}} (r^2+a^2,0,0,a); \qq e_{r}^{(\mu)}=\sqrt{\tfrac{\Delta}{\Sigma}}(0,1,0,0) \q\notag\\
e_{\theta}^{(\mu)}&=&\tfrac{1}{\sqrt{\Sigma}}(0,0,1,0);\q\;\;
e_{\phi}^{(\mu)}=\tfrac{-1}{\sqrt{\Sigma} \sin \theta}(a\sin^2 \theta,0,0,1)\notag
\end{eqnarray}
From the definition $q^{\mu\nu}v_{\mu}v_{\nu}=-1$ this allows one to identify $v^t=\tfrac{r^2+a^2}{\sqrt{\Sigma \Delta}}$ and $\omega=\tfrac{a}{r^2+a^2}$, which implies that the expression of the angular velocity is completely general for every fluid able to support the metric (\ref{eq:rotmetric}). Similarly, the unit space-like vector,  $q^{\mu\nu}\xi_{\mu}\xi_{\nu}=1$, is given by $\xi^\mu=(0,\tfrac{1}{\sqrt{q_{rr}}},0,0)=\tfrac{\Delta}{\Sigma}(0,1,0,0)$. Next, by lowering the indices with the metric (\ref{eq:rotmetric}) and after some algebra one gets the expressions
\begin{equation} \label{eq:unitvec}
v_{\mu}=\sqrt{\tfrac{\Delta}{\Sigma}}(-1,0,0,a\sin^2 \theta)\hspace{0.1cm};\hspace{0.3cm} \xi_{\mu}=\sqrt{\tfrac{\Sigma}{\Delta}}(0,1,0,0)  \ .
\end{equation}
Therefore, the axially symmetric EiBI metric can be found from the general expression (\ref{eq:ggen}) by writing it formally as
\begin{equation} \label{eq:gsol}
g_{\mu\nu}=q_{\mu\nu} + \epsilon \kappa^2 h_{\mu\nu} \ ,
\end{equation}
where  $q_{\mu\nu}$ is the seed GR metric (\ref{eq:rotmetric}), while the EiBI-induced correction $h_{\mu\nu}$ is written as
\begin{equation}\label{eq:hmn}
h_{\mu\nu}\!=\!-\!\left[\tfrac{1}{2}(\rho^q-p_r^q)q_{\mu\nu}+(\rho^q+p_{\perp}^q)v_\mu v_\nu + (p_r^q-p_{\perp}^q) \xi_\mu \xi_\nu\right]
\end{equation}
In this expression the unit vectors are those appearing in (\ref{eq:unitvec}), while the components of the energy-momentum tensor are found via projection on the Carter's tetrad above and comparison with the Einstein equations (\ref{eq:Einfra}). This yields the expressions
\begin{equation} \label{eq:enpre}
\rho^q=-p^q_r=\frac{2\eta' r^2}{\k^2 \S^2} \hspace{0.1cm} ; \hspace{0.4cm} p^q_{\perp}= p^q_r-\frac{\eta'' r + 2\eta'}{\kappa^2 \Sigma} \ ,
\end{equation}
where primes denote derivatives with respect to the radial coordinate $r$. Since both $\rho^q$ and $p_{\perp}^q$ are functions of such a coordinate, the above expressions allow, in principle, to write the tangential pressure as a function of the energy density, {\it i.e.} $p_{\perp}^q=K(\rho^q)$. Then, the EiBI correction (\ref{eq:hmn})  simplifies down to
\begin{equation} \label{eq:hmunu}
h_{\mu\nu}=-\left[\rho^q q_{\mu\nu} +(\rho^q+K(\rho^q))(v_\mu v_\nu-\xi_\mu \xi_\nu)\right]
\end{equation}
Making explicit the unit vectors obtained in (\ref{eq:unitvec}), the components of this correction term can be computed as
\begin{eqnarray}
h_{tt}\,&=&-\tfrac{1}{\Sigma}(\rho^q a^2 \sin^2 \theta + K(\rho^q) \Delta) \label{h1}\\
h_{rr}&=&\tfrac{\Sigma}{\Delta} K(\rho^q) \\
h_{\theta\theta}&=&-\Sigma\,\rho^q  \\
h_{t\phi}&=&\tfrac{a\sin^2\theta}{\Sigma} \left(\rho^q(r^2+a^2)+K(\rho^q) \Delta\right) \\
h_{\phi\phi}&=&-\tfrac{\sin^2\theta}{\Sigma} \left({\rho^q(r^2+a^2)^2+a^2 \sin^2 \theta K(\rho^q) \Delta} \right)\q\label{h5}
\end{eqnarray}

Since the EiBI parameter is heavily constrained by astrophysical \cite{Avelino:2012ge}, particle physics \cite{Delhom:2019wir,Jimenez:2021sqd}, and cosmological \cite{Benisty:2021laq} observations, the metric (\ref{eq:gsol}) will typically involve mild deviations with respect to the GR solutions at the near-horizon/photon sphere scale, at least for astrophysical-size black holes (and even smaller for supermassive ones),  therefore posing a challenge for its observational detectability via gravitational waves and/or shadows. However, strong deviations will indeed appear as getting closer and closer to the innermost region of the solutions, where the EiBI-induced corrections can be the dominant contribution. This fact may give rise to causal and singularity structures remarkably different from those appearing in GR.  Under the presence of horizons these modifications to the causal structure would be barely noticeable from the outside, while horizonless compact objects such as traversable wormholes, which could also arise out of these solutions (see e.g. \cite{Olmo:2013gqa,Afonso:2019fzv} for examples in the spherically symmetric case), would represent much better prospects from an observational point of view.

For further concreteness, we will focus on a specific class of anisotropic fluid matter sources - non-linear electrodynamics (NED) - for which the mapping procedure is remarkable simple, while at the same time allows to find physically relevant solutions.  NEDs are given by Lagrangian densities that, in the Einstein frame, have the form\footnote{Here we neglect magnetic fields for simplicity, though the whole mapping procedure works equally fine should we have included them.} $\tilde{\mathcal{L}}_m=\tilde{\varphi}(Z)$, with $Z=-\tfrac{1}{2}F_{\mu\nu}F^{\mu\nu}$, where $F_{\mu\nu}=\partial_{\mu}A_{\nu}-\partial_{\nu}A_{\mu}$ is the field strength tensor of the vector potential $A_{\mu}$, and $F^{\mu\nu}=q^{\mu\alpha}q^{\nu\beta}F_{\alpha\beta}$. The corresponding field equations, $\nabla_{\mu}(\sqrt{-q} \tilde{\varphi}_Z F^{\mu\nu})=0$, allow to find $Z=Z(\rho^q)$,  so one can see the relation $p^q_{\perp}=K(\rho^q)$ as characterizing the particular NED model under consideration. Furthermore, the interest in these models comes from two different niceties of them.

Firstly, the solution for the (electrostatic) spherically symmetric problem for any NED of the form above within GR is known in exact closed form \cite{Diaz-Alonso:2009xkw}:
\begin{equation} \label{eq:fsss}
f(r)=1-\frac{2M}{r}-\frac{1}{r} \int _r^{\infty} R^2T_0^0(R,Q)dR \ ,
\end{equation}
where $T_0^0(r,Q)=\tilde{\varphi}(Z)-2Z\tilde{\varphi}_Z $ is the temporal component of the energy-momentum tensor in such a spherically symmetric scenario, in which the field invariant can be resolved as $Z\tilde{\varphi}_Z^2=Q^2/r^4$. In such a case, the expressions (\ref{eq:enpre}) of the rotating case turn into
\begin{equation} \label{eq:enpreNED}
\rho^q=-p_r^q=-\frac{r^4T_0^0}{\Sigma^2} \hspace{0.1cm} ; \hspace{0.3cm} p_{\perp}^q=-\rho^q+\frac{4r^2T_0^0+r^3(T_0^0)'}{2\Sigma} \ ,
\end{equation}
which can be in turn interpreted as the energy density and pressures generated by an axially symmetric electromagnetic field with components $A_{\mu}=(A_t,0,0,A_{\phi})$.

Secondly, the mapping equations (\ref{eq:mapp1})-(\ref{eq:mapp3}) provide the correspondence of NEDs in the RBG and GR frames, where the former is characterized by a new function
$\mathcal{L}_m=\varphi(X)$, with $X=-\tfrac{1}{2}B_{\mu\nu}B^{\mu\nu}$, where $B_{\mu\nu}=F_{\mu\nu}$ but $B^{\mu\nu}=g^{\mu\alpha}g^{\nu\beta}F_{\alpha\beta}$, {\it i.e.}, its indices are raised with $g^{\mu\nu}$ rather than with $q^{\mu\nu}$. The explicit shape of the correspondence was worked out in detail in \cite{Delhom:2019zrb}, finding the specific relation between the NEDs on each frame. Therefore, once one defines the target NED in the RBG (or GR) frame, it is an entertaining algebraic gymnastics to find its counterpart on the other frame. We can now use this powerful result in order to work out the counterpart of the Kerr-Newman black hole within EiBI gravity. This involves the coupling of EiBI to Maxwell electrodynamics in such a way that, according to the results of \cite{Delhom:2019zrb},  the corresponding NED on the GR side is given by (again we restrict ourselves to purely electric fields):
\begin{equation} \label{eq:BIem}
\varphi(X)=2\beta^2 \left(1-\sqrt{1-\frac{X}{\beta^2}}\right) \ ,
\end{equation}
with the identification of constants
\begin{equation} \label{eq:beep}
\beta^2=-\frac{1}{2\epsilon \kappa^2} \ .
\end{equation}
This is actually the celebrated Born-Infeld (BI) electrodynamics\footnote{Should one have considered magnetic fields, not only $X$ and $Z$ would have picked a term on them, but also the field invariant associated to the field strength dual $B_{\mu\nu}^{*}=-\tfrac{1}{2}\epsilon_{\mu\nu\alpha\beta}B^{\alpha\beta}$ would pop up in this expression under the desired BI form in (\ref{eq:BIem}). Therefore, this correspondence is exact for all electromagnetic configurations. See \cite{Delhom:2019zrb} for details.}, provided that we stick
to the identification between the BI parameter $\beta$ and the EiBI one in the branch $\epsilon <0$, as forced by (\ref{eq:beep}).
As is well known, for  electrostatic, spherically symmetric solutions, the BI electrodynamics yields a field invariant $X=\tfrac{\beta^2 Q^2}{\beta^2 r^4+Q^2}$ allowing to remove the divergence due to the self-energy of the electron, but not the central black hole singularity when coupled to GR, see {\it e.g.} \cite{Diaz-Alonso:2009xkw}.   However, when moving to the solutions of its EiBI gravity + Maxwell counterpart (in the non-rotating regime), which were discussed in detail in \cite{Afonso:2018mxn}, one finds the geodesic completeness and singularity-free character of the corresponding space-times \cite{Olmo:2015dba}.

In order to build now the rotating version of the EiBI + Maxwell system using the formalism developed above we first need to compute the following quantity, which appears in the GR frame via Eq.(\ref{eq:fsss}):
\begin{equation}  \label{eq:T00}
T_0^0=\frac{2\beta}{r^2}\left( \beta r^2- \sqrt{\beta^2 r^4+Q^2} \right) \ ,
\end{equation}
which reduces to the result of Maxwell electrodynamics on such a frame in the limit $\beta \to \infty$, {\it i.e.}, $T_0^0 \sim Q^2/r^4 + \mathcal{O}(r^{-8})$. However, the above limit cannot be directly taken since, via the identification (\ref{eq:beep}), this would imply taking $\epsilon \to 0$ and EiBI gravity would boil down to GR, thus simply recovering the usual Kerr-Newman solution. Incidentally one can note that should one try to map the GR + Maxwell system ({\it i.e.} the usual Kerr-Newman solution), the mapping would provide the counterpart on the EiBI side as given again by BI electrodynamics, but now the identification of  EiBI/BI constants above holds true only in the $\epsilon >0$ branch. The rotating solutions of such a theory via the mapping were found and discussed in detail in Ref. \cite{Guerrero:2020azx}. Multicenter solutions were also constructed in that way in \cite{Olmo:2020fnk}.

In the present EiBI + Maxwell system under consideration, one just needs to insert the expression (\ref{eq:T00}) into those of the energy density and pressures of the rotating case, Eq.(\ref{eq:enpreNED}), to get
\begin{eqnarray}
\rho^q&=&\frac{2\beta r^2}{\Sigma^2} \left(\sqrt{\beta^2 r^4+Q^2} - \beta r^2\right)  \label{eq:edBI} \\
K&=& -\rho^q+\frac{2\beta}{\Sigma}\left(2\beta r^2-\frac{2\beta^2 r^4+Q^2}{\sqrt{\beta^2 r^4+Q^2}}\right) \ ,  \label{eq:prBI}
\end{eqnarray}
which, as expected, reduce to their non-rotating EiBI-Maxwell counterparts in the limit $a \to 0$ (where $\Sigma \to r^2)$.

We now have everything needed in order to generate the counterpart of the Kerr-Newman solution in the EiBI+Maxwell system via Eq.(\ref{eq:gsol}). The background metric $q_{\mu\nu}$ is given by the line element (\ref{eq:rotmetric}), where the spherically symmetric function $f(r)$ feeding it via (\ref{eq:eta}) is defined by Eq.(\ref{eq:fsss}), after inserting there the expression (\ref{eq:T00}) for the Born-Infeld electrostatic field. The corresponding expression admits an analytic integration to yield the result
\begin{eqnarray}
&&f(r)=1-\frac{2M}{r}  \\
&&+\frac{2 \beta^2}{3} \left[r^2-\sqrt{r^4+\frac{Q^2}{\beta^2}}+\frac{2Q^2}{\beta^2 r^2} \ _{2}F_1\left[\frac{1}{4},\frac{1}{2},\frac{5}{4},\frac{-Q^2}{\beta^2 r^4} \right] \right] \nonumber
\end{eqnarray}
where $_2F_1[a,b,c;z]$ is a hypergeometric function. This unambiguously defines the shape of the seed (GR-based) rotating solution, parameterized by mass $M$, charge $Q$, spin $a$, and Born-Infeld constant $\beta$. As for the (EiBI-driven) correcting term $h_{\mu\nu}$, its components are given in Eqs.(\ref{h1})-\eqref{h5}, after inserting the expressions (\ref {eq:edBI}) and (\ref {eq:prBI}) for the energy density and pressure of the rotating fluid, respectively. This completes our construction of the axially symmetric metric of the EiBI + Maxwell system, {\it i.e.}, the counterpart of the Kerr-Newman metric.

Note that in this Kerr-Newman-like metric the BI-dependencies in $\beta^2$ are replaced by EiBI dependencies in $\epsilon$ via the identification (\ref{eq:beep}), which therefore becomes the only parameter encoding the deviations of this rotating metric with respect to the Kerr-Newman solution of GR.

It should also be noted that far from the sources, $r \to \infty$, where the energy density and pressure approximate their vacuum values, $\rho^q \approx p_{\perp}^q \approx 0$, one has $g_{\mu\nu} \approx q_{\mu\nu}$, and the EiBI Kerr-Newman line element  boils down to
\begin{eqnarray}
ds_g^2&=&- \left(1-\frac{2M}{r}+\mathcal{O}(r^{-2})\right) dt^2 \nonumber \\
&+& \left(1+\mathcal{O}(r^{-1})\right)(dr^2+r^2 d\Omega^2) \\
&+&\left(\frac{4a M\sin^2 \theta}{r} + \mathcal{O}(r^{-2})\right)dtd\phi \ , \nonumber
\end{eqnarray}
in agreement with the expectations of the weak-field limit of the Kerr-Newman solution of GR. The EiBI corrections appear at order $\mathcal{O}(r^{-6})$ and therefore they are strongly suppressed in this limit. As a consequence, this EiBI Kerr-Newman metric naturally passes the same weak-field limit tests as the GR Kerr-Newman does, such as the consistence with star's orbital motions around them. However, in the strong-field regime the EiBI corrections in the parameter $\epsilon$, fueled by the growth of the energy density and pressure appearing in the $h_{\mu\nu}$ term of Eq.(\ref{eq:gsol}), become non-negligible, to the point that in the innermost region of the solution it may constitute the dominant contribution.

Due to the fact that, in particular, the $g_{tt}$ and $g_{rr}$ components in the EiBI Kerr-Newman solution get corrections in the energy density and pressure (suppressed by the EiBI length-squared scale), modifications to the astrophysically relevant surfaces of their GR counterparts do occur. Indeed, it affects both the (ergo-)horizons and the three-dimensional shell
of unstable null geodesics \cite{Johnson:2019ljv}.
Since these are the main geometrical features needed for the analysis of gravitational wave radiation \cite{Ezquiaga:2020dao} as well as shadows and light rings from accretion disks \cite{Gralla:2019xty,Chael:2021rjo} their study is of uttermost interest. On the other hand, the singularity structure in terms of geodesic completeness and behaviour of curvature scalars is also modified. For instance, the ring singularity of the GR Kerr-Newman solution, which is revealed in the zeros of the $g_{\theta\theta}=\Sigma$ component, and amounts to $\{r=0,\theta=\pi/2\}$, is modified in the present EiBI Kerr-Newman geometry to $\Sigma(1-\epsilon \kappa^2 \rho^q)=0$, where the energy-density corrections appearing via (\ref{eq:edBI}) are again the relevant protagonist. An in-depth analysis of these features will be carried out elsewhere.

\section{Conclusion}

In this work we have combined several theoretical results found in the last few years in order to produce an infinite class of exact axisymmetric solutions of modified gravity, which are suitable frameworks for representing alternative black holes beyond the Kerr-Newman solution.  Using anisotropic fluids as the matter source, this has been achieved by employing a  mapping  between the spaces of solutions of certain families of metric-affine gravities, dubbed as Ricci-based gravities, and GR itself. This mapping method allows to obtain a solution of the former using a seed solution of the latter via purely algebraic transformations, once the identification between the matter Lagrangian densities on each side of the mapping is made.

In order to obtain this class of axisymmetric solutions we have first considered the Newman-Janis technique which, starting from a general spherically symmetric metric, allows to find its rotating counterpart, assuming GR as the  theory of gravity supporting it. At the same time, the use of the Carter's tetrad allows to find the energy density, pressure, and angular velocity of the fluid generating such an axisymmetric solution by demanding the fulfillment of Einstein equations. This general family of GR rotating solutions thus becomes the seed on which to grow the map. In the case of EiBI gravity, a particularly agreeable member of the RBG class, this allows to find its rotating counterpart in exact closed form, which can be decomposed as the original GR one plus a correction term(s) suppressed by the EiBI length-squared scale. Next we turned to (non-linear) electromagnetic fields, which can be described as anisotropic fluids where the choice of the relation $p_{\perp}^q=K(\rho^q)$ characterizes a particular model. Using the fact that the corresponding spherically symmetric solutions (to be Newman-Janis-transformed) are known in exact form for any such model, we first found the correspondence between EiBI gravity coupled to a Maxwell field and GR coupled to a Born-Infeld-type electrodynamics, and next combined both results to determine the counterpart of the Kerr-Newman solution in this framework. This identification requires the EiBI constant to be negative, which is precisely the branch of solutions that, in the spherically symmetric case, removes the singularity issue via a restoration of geodesic completeness for every null and time-like curve.

In our opinion, this modified Kerr-Newman solution of the EiBI + Maxwell system, which illustrates the capabilities of the mapping method, is the most succulent result of the present work. While it reduces in the weak-field limit to the GR Kerr-Newman one, therefore being automatically compatible with the observed orbital motions in such a regime, it encapsulates the departures from it arising in the strong-field regime via a single additional (EiBI) parameter. For non-vanishing values of the energy density of the matter (electromagnetic) field, the modified gravitational physics yields departures from GR predictions in terms of (ergo-)horizons, critical (null) surfaces, and the innermost structure of the black hole, the latter affecting the ring-like singularity issue. These aspects are of great interest not only from a purely theoretical point of view, but also for the sake of multimessenger astronomy. This is so because EiBI-induced effects  happening outside the event horizon are expected to yield modifications to the waveforms of gravitational waves out of binary mergers, and to the shape of their optical appearance and shadows when illuminated by accretion disks. Besides, should the successful removal of space-time singularities of the spherically symmetric EiBI+Maxwell system be extended to the axisymmetric case, this would imply that horizonless Kerr-Newman solutions of this kind  (supplemented with a suitable gravitational collapse mechanism) would also be acceptable as compact objects  alternative to GR black holes, {\it i.e.}, black hole mimickers \cite{Shao:2020weq}. Combined, these results would allow to test the existence of new gravitational dynamics beyond the present weak-field limit tests \cite{EventHorizonTelescope:2020qrl}. We are currently working to extract and analyze all these features from the above solutions and we hope to report on this issue very soon.

\section*{Acknowledgments}

DRG is funded by the \emph{Atracci\'on de Talento Investigador} programme of the Comunidad de Madrid (Spain) No. 2018-T1/TIC-10431. This work is supported by the Spanish Grants FIS2017-84440-C2-1-P, PID2019-108485GB-I00 and PID2020-116567GB-C21 funded by MCIN/AEI/10.13039/501100011033 (``ERDF A way of making Europe" and ``PGC Generaci\'on de Conocimiento"), the project PROMETEO/2020/079 (Generalitat Valenciana), the project H2020-MSCA-RISE-2017 Grant FunFiCO- 777740, the project i-COOPB20462 (CSIC),  the FCT projects No. PTDC/FIS-PAR/31938/2017 and PTDC/FIS-OUT/29048/2017, and the Edital 006/2018 PRONEX (FAPESQ-PB/CNPQ, Brazil, Grant 0015/2019).  VIA thanks the Theoretical Physics Department of the Complutense University of Madrid for kind hospitality while doing this work. This article is based upon work from COST Action CA18108, supported by COST (European Cooperation in Science and Technology).

\end{document}